\documentclass[conference]{IEEEtran}

\usepackage{amssymb}
\usepackage[cmex10]{amsmath}
\usepackage[utf8]{inputenc}
\usepackage{graphicx}
\usepackage{subfigure}

\usepackage{hbyMacros}

\title{Effect of ISI Mitigation on Modulation Techniques in Communication via Diffusion}

\author{H. Birkan Yilmaz, Na-Rae Kim, Chan-Byoung Chae \\
School of Integrated Technology, Yonsei University, Korea\\
E-mail:\{birkan.yilmaz, nrkim, cbchae\}@yonsei.ac.kr}

\begin{document}

\maketitle
\begin{abstract}
Communication via diffusion (CvD) is an effective and energy efficient method for transmitting information in nanonetworks. In this work, we focus on a diffusion-based communication system where the reception process is an absorption via receptors. Whenever a molecule hits to the receiver it is removed from the environment. This kind of reception process is called first passage process and it is more complicated compared to diffusion process only. In 3-D environments, obtaining analytical solution for hitting time distribution for realistic cases is complicated, hence we develop an end-to-end simulator for he diffusion-based communication system that sends consecutive symbols. 

In CvD, each symbol is modulated and demodulated in a time slot called symbol duration, however the long tail distribution of hitting time is the main challenge that affects the symbol detection error. The molecules arriving in the following slots become an interference source when detection takes place. End-to-end simulator enables us to analyze the effect of inter symbol interference (ISI) without making any assumptions on the ISI. 
%Due to differences in channel response with respect to electromagnetic case, %conventional ISI mitigation techniques cannot be applied directly. 
We propose an ISI cancellation technique that utilizes decision feedback for compensating the effect of previously demodulated symbol. Three different modulation types are considered with pulse, square, and cosine carrier waves. In case of constraints on transmitter or receiver node it may not be possible to use pulse as a carrier, and peak-to-average messenger molecule metric is defined for this purpose. Results show that, the proposed ISI mitigation technique improves the symbol detection performance and the amplitude-based modulations are improved more than frequency-based modulations. 
\end{abstract}


\section{Introduction}

Nanonetworking is to interconnect several nanoscale machines that has molecular communication as one of the most promising mechanisms~\cite{aky08,atakan2007information,Akan_ban12,far11, hiy08,hiy05,nak07,nak05}. In molecular communication, nanoscale machines or nanonetworking-enabled nodes communicate each other by exchanging information via chemical/biological molecules which is called messenger molecules.
The messenger molecules carrying information propagate from a transmitter to a receiver nanonetworking enabled node by walking along a specific rail, diffusion, or drift in a certain fluid medium~\cite{direct13,far11,TNBS_Kim13,mahfuz2010characterization,srinivas2012molecular}. 
Depending on each propagation scheme, theoretical analysis such as calculating channel capacity can be achieved through information theoretic approach. Practical analysis becomes also possible with the first molecular communication platform developed by the authors in~\cite{far13}. Realistic channel and noise models are also developed from channel measurements using a molecular communication testbed platform~\cite{jsac13}.

Among the propagation schemes, molecular communication driven by diffusion has been most profoundly researched, and it is often called the Communication via Diffusion (CvD), which is also a main focus of this work~\cite{kuran2010energy,TSP_11}. 
Diffusion process, or Brownian motion, is basically a random process, and the propagation channel is modeled as an inverse Gaussian distribution, accordingly~\cite{srinivas2012molecular}. At the receiver side, randomly propagated messenger molecules are assumed to immediately disappear after arriving, which makes first hitting time an important measurement~\cite{redner2001guide}. If we deal with the first hitting process, considering the concentration formulation, $C(r,t)$, is not the correct way of finding the hitting time histogram. It is hard to find an analytical solution of hitting time distribution in a three dimensional environment, however~\cite{redner2001guide}, though one dimensional analytical solution is already proved. Therefore, a proper three dimensional simulator is required to precisely track the behavior of messenger molecules within a realistic environment. In the literature, several simulators are proposed in~\cite{gul2010nanons, llatser2011n3sim} but the reception process at the receiver site is not well implemented. Rather, the simulators mostly focus on the diffusion process with assuming that the messenger molecules can go through the receiver boundary in and out, and also they focus on one-shot transmission. 

Another important factor from first hitting time distribution is ISI~\cite{TNBS_Kim13,kuran2012interference,lin2012signal, Noel_ISI13}. A general hitting time distribution of a molecular communication system shows a very long tail in terms of time scale. Specifically, in a certain symbol duration, there exists surplus messenger molecules from the previous symbol as well as transmitted messenger molecules from the current symbol. That means ISI will be accumulated when sending consecutive symbols, and its effect is inevitable unless symbol duration or interval becomes infinite, which is highly inefficient, or there is a strong medium drift~\cite{TNBS_Kim13}. Thus, ISI should be properly eliminated to have a better system performance.  

Thus, in this paper, a three dimensional molecular communication simulator is developed to track all the transmitted messenger molecules, and a special type of filter is applied to mitigate ISI effect. We also define Peak-to-Average Messenger Molecule (PAMR) metric for quantifying the constraints on the communicating pairs. Besides, one novel modulation technique is also proposed that is Molecular Frequency Shift Keying (MFSK). Several other modulation techniques have been addressed from previous work, and there are pros and cons depending on system conditions. For example, information can be encoded in concentration, type, ratio, or release timing of messenger molecules~\cite{JSAC_Kim12,kuran2011modulation}. A new modulation to be proposed here uses a cosine wave of transmission signal with different frequency values for different symbols.  

The paper is organized as follows. Section~\ref{Sec:system} introduces the system model focusing on the CvD processes. New modulation and ISI mitigation techniques are presented in Section~\ref{Sec:modulation_and_isi}. Performance evaluation in Section~\ref{Sec:performance_evaluation} is followed by the conclusion in Section~\ref{Sec:conclusion}.

\section{System Model}
\label{Sec:system}
In this section we present the details of diffusion/reception processes and CvD system.

\subsection{Diffusion and the First Hitting Time}
The messenger molecules are the information particles in
molecular scale. In this scale, the movement of particles inside a fluid is modeled by Brownian motion or diffusion process. The motion is governed by the combined forces applied to the messenger molecule by the molecules of the liquid due to thermal energy. If we just consider the diffusion process of a particle starting from origin then the concentration at radius $r$ and time $t$ is given by the following formula
% % % % % % % % % % % % % % % % % % % % % % % % % % %
\beqn
C(r,t) = \frac{1}{(4 \pi D t)^{n/2}} e^{-r^2/4Dt}
\eeqn
% % % % % % % % % % % % % % % % % % % % % % % % % % % 
where $n$ and $D$ are the dimension of the environment and the diffusion coefficient, respectively \cite{redner2001guide}. We can easily scale the formula to releasing many molecules by multiplying with the number of molecules released. The value of $D$ depends on temperature of the environment, viscosity of the fluid, and the Stokes' radius of the molecule \cite{tyrrell1984diffusion}. 

In nature, whenever a messenger molecule's body coincides with the body of the receiver, the molecule is received and removed from the environment. Therefore, after that point that molecule cannot move further and constitutes the signal just once. This process is named first passage or hitting process and we are concerned with the probability that a diffusing particle first reaches a specified site or sites at a specified time. 

In 1-D environment we have the closed form solution for first hitting probability function, however for the 2-D or 3-D environments even with a  symmetrical receiver, closed form solution is a hard surface integration or differential equation problem. In 1-D and 2-D environments diffusing particles hit the receiver in the long run with probability 1 (recurrent process). When we consider 3-dimensional environment, however there is a nonzero probability for a diffusing particle to miss the receiver \cite{redner2001guide}.

In 1-D environment the first hitting probability for a point source is 
% % % % % % % % % % % % % % % % % % % % % % % % % % % % % % % %
\beqn
F_h(r_0,t) = \frac{r_0}{\sqrt{4 \pi D t^3}} e^{-r_0^2/4Dt}
\eeqn
% % % % % % % % % % % % % % % % % % % % % % % % % % % % % % % %
where $r_0$ is the distance to the absorber point. First hitting probability in 1-D environment is inversely proportional in $t^{3/2}$. We do not have an efficient closed form solution in 3-D environment when the source node is also spherical, hence we run Monte Carlo simulations, at each small time step molecules move according to diffusion dynamics. 

In $n$-dimensional space, the total displacement, $\displacementv$ in one time step can be found as 
% % % % % % % % % % % % % % % % % % % % % % % % % % % % % % %
\beqn
\displacementv = (\Delta x_1, ... , \Delta x_n)
\eeqn
% % % % % % % % % % % % % % % % % % % % % % % % % % % % % % %
where $\Delta x_i$ is the displacement at the $\ith$ dimension. Movement at each dimension in one time step is modeled independently and follows a Gaussian distribution
% % % % % % % % % % % % % % % % % % % % % % % % % % % % % % %
\beqn
\Delta x_i \sim \mathcal{N}(0, 2D\Delta t) \onlygap \forall i \in \{1,...,n\}
\label{eqn:diffusion1}
\eeqn
% % % % % % % % % % % % % % % % % % % % % % % % % % % % % % %
where $\Delta t$ is the step time. Molecules propagate in the environment according to these dynamics. We model transmitter, receiver and the messenger molecules as spherical bodies. Our model ignores collisions between the messenger molecules, as done in the literature for simplicity \cite{moore2009molecular}. The transmitter node is a reflecting boundary for messenger molecules, that makes the closed form solution more harder since it changes the symmetry for the solution. Therefore, we simulate the first hitting process by moving messenger molecules according to (\ref{eqn:diffusion1}) and reflecting or removing the particles that hit to transmitter or receiver, respectively. 
%%%%%%%%%%%%%%%%%%%%%%%%%%%%%%%%%%%%%%%%%%%%%%%%%%%%%%%%%%%%%%%%%%%%%%%%%%%%%%%%
\begin{figure}[!t]
\centering
\includegraphics[width=0.99\columnwidth,keepaspectratio] {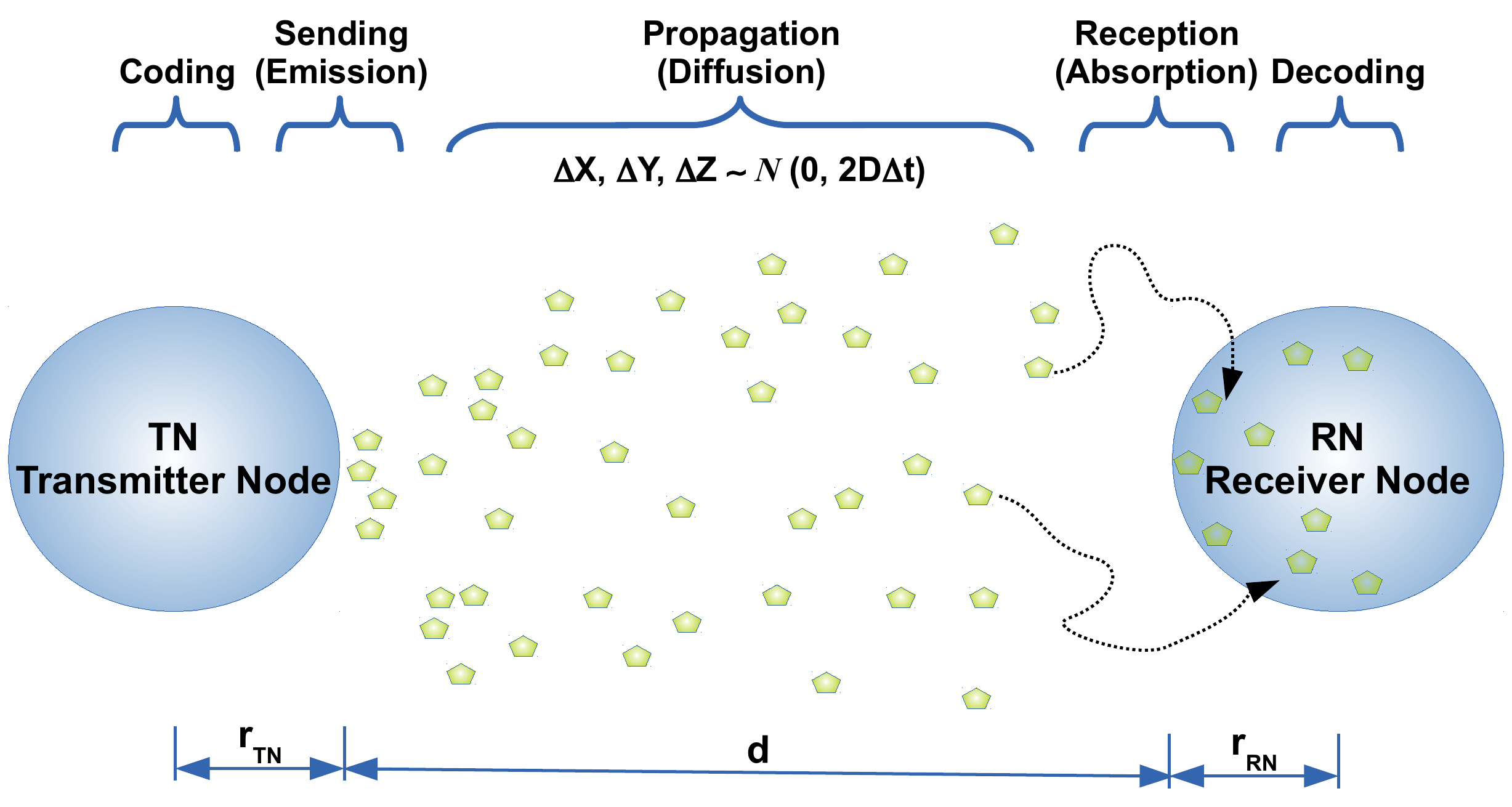}
\caption{CvD system model in 3-D environment.}
\label{fig:cvd_system_process_model}
\end{figure}
%%%%%%%%%%%%%%%%%%%%%%%%%%%%%%%%%%%%%%%%%%%%%%%%%%%%%%%%%%%%%%%%%%%%%%%%%%%%%%%%

\subsection{Communication via Diffusion}
We model a communication system composed of a fluid environment and a pair of devices, each called nanonetworking-enabled node; one as the transmitter and one as the receiver. In CvD, the information is transmitted between the transmitter and receiver through the propagation of certain molecules via diffusion \cite{pierobon2010physical}.  These molecules can be of many types of chemical compounds such as DNA fragments, proteins, peptides or specifically formed molecules as long as they are biocompatible with the inhabitant \cite{kuran2010energy}.

The CvD system for distance $d$ is depicted in Fig.~\ref{fig:cvd_system_process_model}, the radii of transmitter and receiver node are denoted by $\rtn$ and $\rrn$, respectively. CvD system has five main processes: encoding, emission, propagation, absorption, and decoding \cite{kuran2012interference}. Some channel coding techniques may be applied during encoding process before modulating the symbols during the emission process. After emitting the messenger molecules, the characteristics of the diffusion channel play an important role on propagation of the molecules. Some of these messenger molecules arrive at the receiver (i.e., hit the receiver), and they form chemical bonds with the receptors on the surface of the receiver \cite{atakan2007information}. When they hit the receiver they are removed from the channel and the properties of these received molecules (e.g., concentration, type) constitute the received signal. Demodulation function takes place during the absorption process and it is followed by decoding if a channel coding is employed.

In a CvD system communicating pairs are assumed to be synchronized and time is divided into equal duration slots in which a single symbol can be sent. These time slots are called symbol durations and denoted by $t_s$. Information is modulated on some of the physical properties of the messenger molecules, it can be the number, type, and any other property of the arriving messenger molecules. In this study, we mainly focus on emission (modulation), propagation (diffusion dynamics), and absorption (demodulation) processes. We also focus on the ISI mitigation for enhancing signal demodulation. 

\subsection{Emission \& Reception Process}
The task of the particle emission process is to modulate the particle concentration $\sentXt$ at the transmitter according to the input symbol $\sentSymt$, modulation type, and the waveform of the signal. In one $t_s$ duration, the number of released molecules may be spike like at the start of the slot or may be spread over the slot. This behavior depends on the capabilities of the transmitter node, input signal, and the sampling period ($\tss$) of the transmitter node. 

Releasing all the molecules for a symbol may be better compared to spreading it over the symbol slot in terms of symbol demodulation. The capabilities of the transmitter node, however, may necessitate the spreading it over the symbol duration. PAMR at the transmitter node is defined as 
\beqn
\pammrtx = \frac{\max \sentXt}{\avgop \sentXt} \,.
\eeqn
PAMR is a similar concept to Peak-to-Average-Power-Ratio (PAPR) in an OFDM system. If the transmitter node has less space for storing the synthesized messenger molecules it would necessitate sending the synthesized messenger molecules before the storage areas are full. Hence, depending on the capabilities of the transmitter node, sending one peak at the start of the symbol duration may be possible or not. Having high PAMR value may violate the transmitter node constraints due to capabilities.  Similarly, PAMR at the RN side can be defined and determines the capabilities/constraints of the RN.  

On the other hand, reception process is assumed to be perfect and whenever an messenger molecule hits the receiver it is directly absorbed and removed from the environment \footnote{Note that, this assumption may not be feasible in practice, thus the authors in \cite{far13} considered imperfect receivers.}. Receptor heterogeneity and the deployment are out of scope of this paper.

%%%%%%%%%%%%%%%%%%%%%%%%%%%%%%%%%%%%%%%%%%%%%%%%%%%%%%%%%%%%%%%%%%%%%%%%%%%%%%%%
\begin{figure}[!t]
\centering
\includegraphics[width=0.99\columnwidth,keepaspectratio] {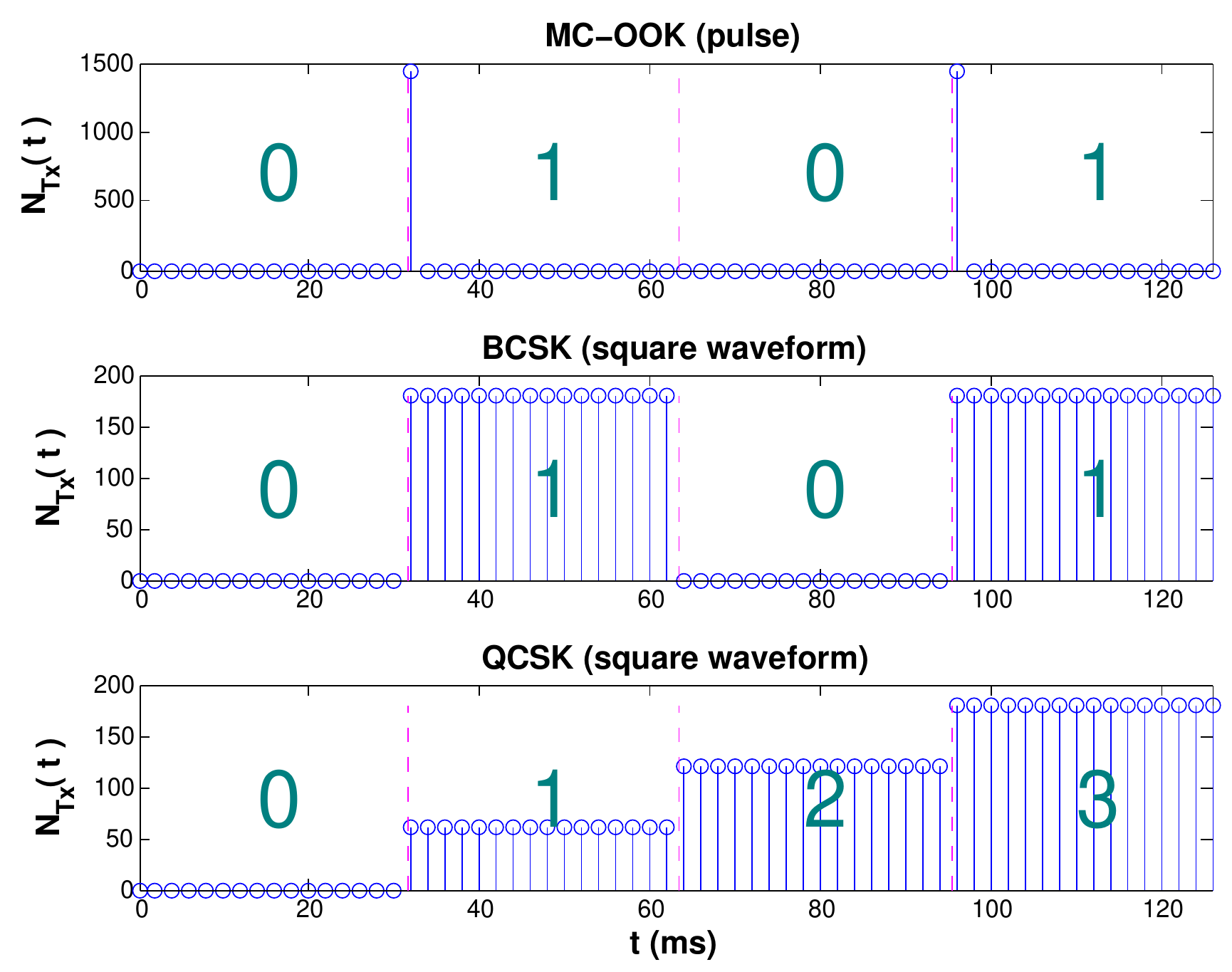}
\caption{$\sentXt$ for amplitude-based modulations.}
\label{fig:amplitude_modulation_waveforms}
\end{figure}
%%%%%%%%%%%%%%%%%%%%%%%%%%%%%%%%%%%%%%%%%%%%%%%%%%%%%%%%%%%%%%%%%%%%%%%%%%%%%%%%
\section{Modulation Techniques and ISI Analysis}
\label{Sec:modulation_and_isi}
In a CvD system, the information is sent using a sequence of symbols spread over sequential time slots as one symbol in each slot. The symbol can be modulated over various physical  properties at the sender, e.g., concentration, frequency, phase, molecule type, to form a signal. Similar properties of the received messenger molecules are used for signal detection and the information is demodulated.

The concentration of the received messenger molecules is used as the amplitude of the signal in Concentration Shift Keying (CSK), hence it represents data as variations in the amplitude of a carrier wave. If the carrier wave is single point pulse and we use presence and absence of carrier wave to indicate binary ``1" and ``0", respectively, it is called Molecular Concentration On-off Keying (MC-OOK). If the carrier wave is a square wave then we call it Binary CSK (BCSK) modulated on a square wave. Similarly, if there are four amplitude levels for two bit symbols then it is called Quadrature CSK (QCSK) modulated on a square wave. These modulations are depicted in Fig.~\ref{fig:amplitude_modulation_waveforms}.

General formula of $\sentXt$ for symbol $s_i$  is denoted by  $\sentXsit$ and $\sentXsit$ with CSK modulations on square waveform is given in (\ref{eqn:receivedSignalCSK}) 
\beqn
\sentXsit =   2 * i * \meanAmp / m \onlygap i=0,...,m
\label{eqn:receivedSignalCSK}
\eeqn
where $s_i$, $m$, and $\meanAmp$ are symbol $i$, maximum symbol id, and mean amplitude, respectively. 

Representing the data through discrete frequency changes of a carrier wave of number of messenger molecules is called Molecular Frequency Shift Keying (MFSK) . In this modulation scheme, number of molecules sent is not constant during the symbol duration and changes according to cosine wave. Since we cannot send negative number of molecules we have to shift the amplitude in y-axis to up. 

\begin{figure}[!t]
\centering
\includegraphics[width=0.99\columnwidth,keepaspectratio] {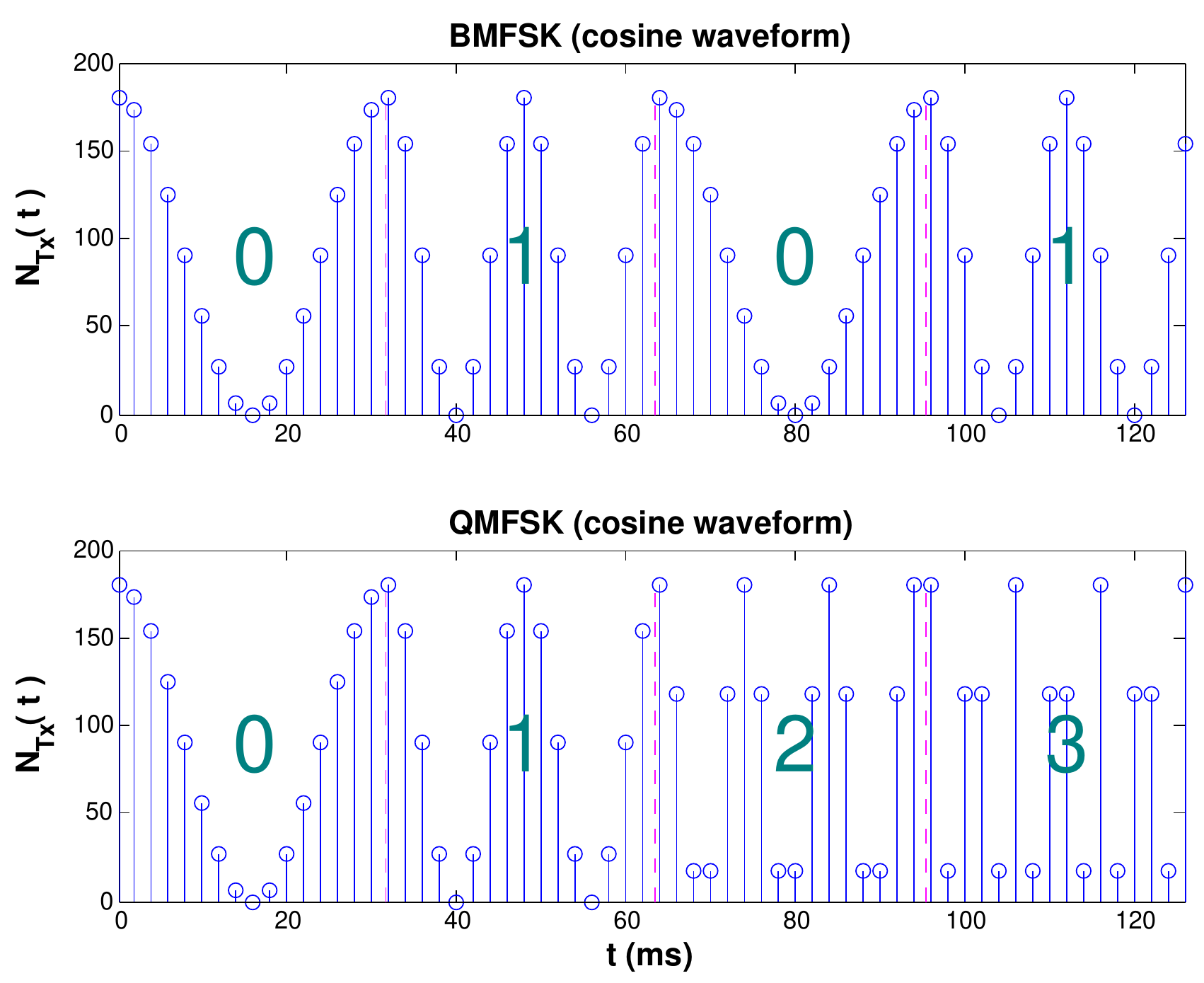}
\caption{$\sentXt$ for frequency-based modulations.}
\label{fig:freq_modulation_waveforms}
\end{figure}

If the carrier wave is a cosine wave and there are two frequencies then it is called Binary MFSK (BMFSK). Similarly, if there are four frequencies for two bit symbols then it is called Quadrature MFSK (QMFSK). These modulations are illustrated in Fig.~\ref{fig:freq_modulation_waveforms}. 
%%%%%%%%%%%%%%%%%%%%%%%%%%%%%%%%%%%%%%%%%%%%%%%%%%%%%%%%%%%%%%%%%%%%%%%%%%%%%%%%
%%%%%%%%%%%%%%%%%%%%%%%%%%%%%%%%%%%%%%%%%%%%%%%%%%%%%%%%%%%%%%%%%%%%%%%%%%%%%%%%
 
General formula for $\sentXsit$ with MFSK modulations on cosine waveform is as follows 
\beqn
\sentXsit =   \meanAmp  + \meanAmp * \cos (2 \pi f_i t) \onlygap i=0,...,m
\eeqn
where $s_i$, $f_i$ and $\meanAmp$ are symbol $i$, frequency for $s_i$, and mean amplitude, respectively. 

If the data is modulated on the amplitude variations at the receiver side a thresholding is employed for detecting symbols. For example, demodulating MC-OOK or BCSK symbols one threshold is used. The receiver decodes the intended symbol as ``1" if the number of messenger molecules arriving at the receiver during a time slot exceeds a threshold $\threshold$, ``0" otherwise. The CvD system using CSK-based modulation can be affected adversely from Inter Symbol Interference (ISI) which can be caused by the surplus molecules from the previous symbols. Due to the diffusion dynamics, some messenger molecules may arrive after their intended time slot. These molecules cause the receiver to decode the next intended symbol incorrectly. 

If the data is modulated on the frequency variations, received symbol at the receiver side can be detected via correlating the received signal with the symbol waveforms after synchronizing the signals. You can also look the received signal at the frequency domain and correlate to the symbol frequency signatures. The CvD system using MFSK-based modulation is less affected from the ISI, however, CSK-based modulations are easier to detect. Symbol detector function $\delta(.)$ for MFSK-based modulations is given in (\ref{eqn:detectorMFSK}).
%%%%%%%%%%%%%%%%%%%%%%%%%%%%%%%%%%%%%%%%%%%%%%%%%%%%%%%%%%%%%%%%%%%%
\beqn
\delta(\rcvYslot{t_0}) = \arg\max\limits_i \suml_{t=t_0:\tss }^{t_0+\ts}\rcvYt \sentXsit 
\label{eqn:detectorMFSK}
\eeqn
%%%%%%%%%%%%%%%%%%%%%%%%%%%%%%%%%%%%%%%%%%%%%%%%%%%%%%%%%%%%%%%%%%%%
where $\rcvYslot{t_0}$ denotes the received signal vector in the symbol duration starting at $t_0$ and the summation advances with sampling duration $\tss$.

\subsection{ISI Mitigation}
We assume that, in each symbol duration, only one symbol is sent. Arising from the probabilistic dynamics of Brownian motion, the messenger molecules move randomly and do not necessarily reach the RN. Instead, every messenger molecule has a probability of hitting the receiver in a predetermined time duration $\ts$ when they are released at during that time duration. The received molecules form the received signal and ISI due to surplus molecules from the previous symbols affects the symbol decoding (demodulation) process severely in CvD since the arrival of messenger molecules spreads to a very long duration. In stead of having longer symbol interval we propose to have a ISI mitigation. 

%%%%%%%%%%%%%%%%%%%%%%%%%%%%%%%%%%%%%%%%%%%%%%%%%%%%%%%%%%%%%%%%%%%%%%%%%%%%%%%%
\begin{figure}[!t]
\centering
\includegraphics[width=0.99\columnwidth,keepaspectratio] {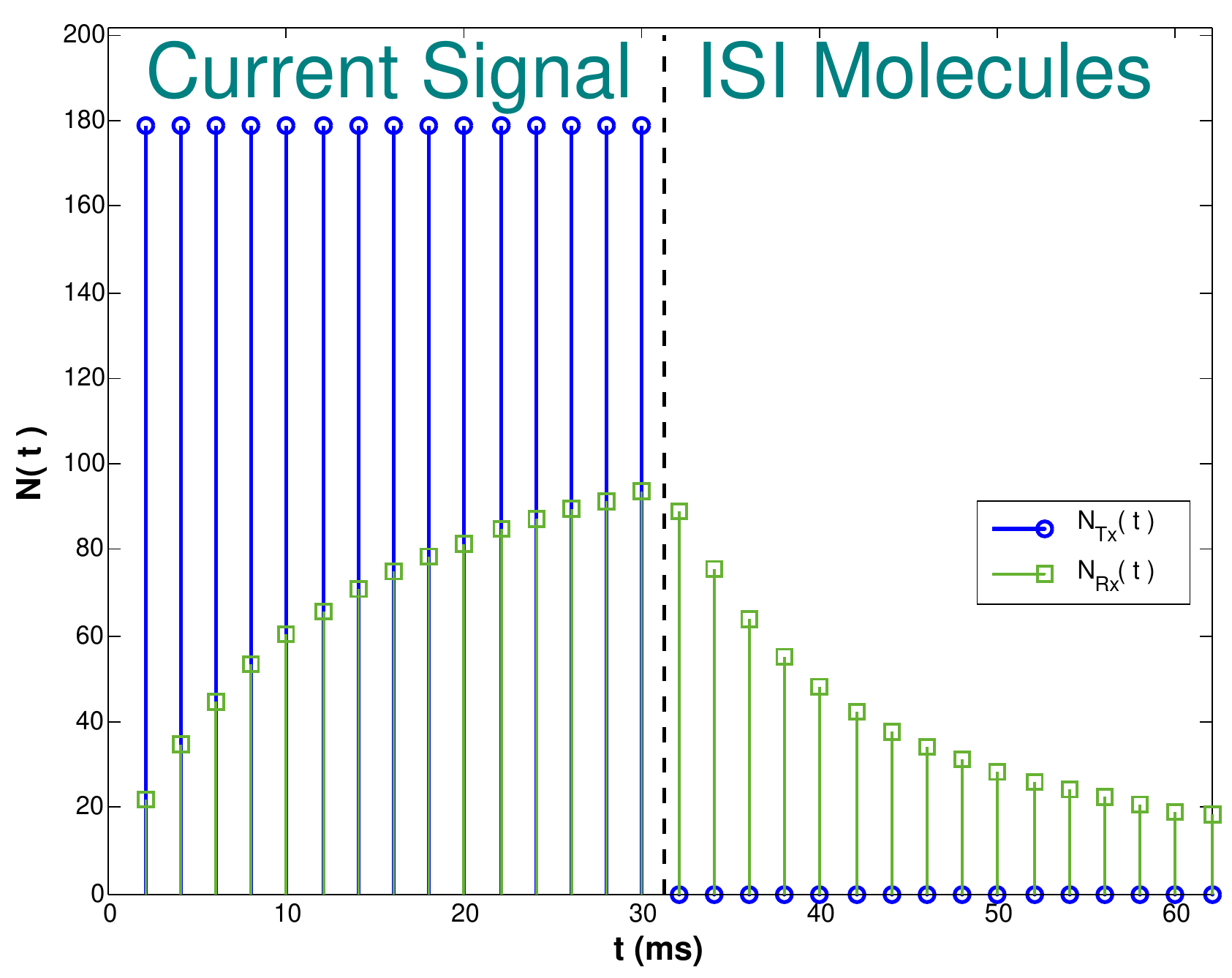}
\caption{ISI example for CSK-based symbol.}
\label{fig:isi_example_freq}
\end{figure}
%%%%%%%%%%%%%%%%%%%%%%%%%%%%%%%%%%%%%%%%%%%%%%%%%%%%%%%%%%%%%%%%%%%%%%%%%%%%%%%%
In prior work it is shown that the current symbol is mostly affected by the previous symbol \cite{kuran2012interference}. Hence, we employ a vector-based  2-tap Decision Feedback Equalizer (DFE) filter at the receiver side, just considering the previous symbol detection decision and canceling its effect on the current symbol. Output signal is modeled as
% % % % % % % % % % % % % % % % % % % % % % % % % % % % % % % % % % % %
\beqn
\rcvYslot{t} = H_0 \sentXslot{t} + H_1 \sentXpslot{t}
\label{eqn:signal_tap_model}
\eeqn
% % % % % % % % % % % % % % % % % % % % % % % % % % % % % % % % % % % %
where $H_k$, $\rcvYslot{t}$ and $\sentXslot{t}$ denote the weighting diagonal matrices, received and sent  signal vector during a symbol duration starting from $t$, respectively. Since we have signal vectors, the weighing coefficients are square matrices of dimension $\ts /\tss$.  $\rcvYslot{t}$ can be seen as a sum of two received signal vectors, one is due to current sent symbol and the other is due to previous symbol. These two components of received signal vector are estimated via extensive Monte Carlo simulations depending on the distance ($d$), messenger molecule, transmitter and receiver node radii ($\rmm$,  $\rtn$, $\rrn$), diffusion coefficient ($D$), symbol duration ($\ts$), and sampling duration ($\tss$). 

%%%%%%%%%%%%%%%%%%%%%%%%%%%%%%%%%%%%%%%%%%%%%%%%%%%%%%%%%%%%%%%%%%%%%%%%%%%%%%%%
\begin{figure*}[!tb]
\centering
\includegraphics[width=1.89\columnwidth,keepaspectratio] {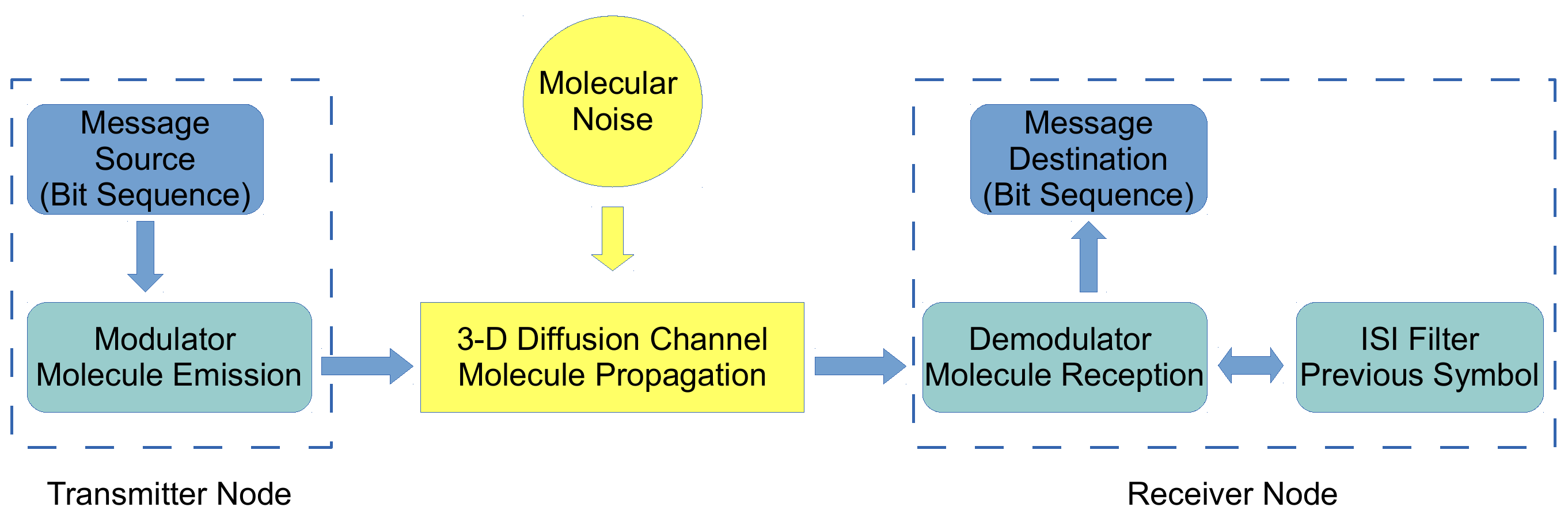}
\caption{End-to-end block diagram of CvD simulator.}
\label{fig:simulator_block_diagram}
\end{figure*}
%%%%%%%%%%%%%%%%%%%%%%%%%%%%%%%%%%%%%%%%%%%%%%%%%%%%%%%%%%%%%%%%%%%%%%%%%%%%%%%%
In Fig.~ \ref{fig:isi_example_freq}, $\rcvYt$ example when symbols are modulated according to BMFSK and sent at the first symbol duration is depicted. Two different symbols with different $f_i$ have different received signal signature depending on the frequency of emission and the waveform. In the first symbol duration received molecule count is shown and in the second symbol duration surplus molecules causing ISI is shown. Circle and square top signals belong to different frequencies, square top symbols' frequency is higher. We acquire these signal signatures via extensive Monte Carlo simulations, and we remove the effect of previous symbol by subtracting the received signal vector due to previous symbol. In MFSK-based modulations the detector uses correlation function, hence we also employ simple synchronization at the receiver side for increasing the effectiveness of the symbol detector.

\subsection{Simulator Model}
To the best of our knowledge, our CvD simulator is the first end-to-end simulator that has modulation, demodulation, and ISI cancellation modules and it considers absorption process for the signal reception in 3-D environment. In this simulator, all the information symbols are sent consecutively without just focusing on one symbol duration. Hence, all the remaining molecules cause ISI for the following symbols. In the literature the ISI effect is not considered adequately, in stead just one symbol ISI is assumed and some semi-analytical formulations are derived \cite{kuran2012interference}. In \cite{lin2012signal}, 1-D environment is considered to derive the analytical formulations for ISI mitigation. We need to analyze the ISI effect in 3-D environment and send the symbols consecutively without making assumptions to have more realistic results. Therefore, having an end-to-end simulator enables us to see the effect of ISI completely in 3-D environment. Entities and the flow diagram of the simulator is depicted in Fig.~\ref{fig:simulator_block_diagram}.

Simulator entities are transmitter and receiver node, messenger molecules, and molecular diffusion channel. The transmitter node has a message source and it randomly creates $\numsym$ symbols to transmit and gives these symbols to modulator module. Modulator module determines the number of molecules to emit ($\sentXt$) at a given time according to chosen modulation scheme and the current symbol. Emitted molecules moves according to 3-D Brownian motion and some of the molecules hit the receiver and they are absorbed by the receptors and removed from the environment. In each symbol duration these steps are repeated consecutively and none of the molecules are removed from the environment if they do not hit the receiver. In each symbol duration, sent and the received molecules are sampled every $\tss$ milliseconds. RN node receives two signals from the channel one is with molecular noise is added and the other one is without noise for just analyzing the diffusion dynamics. At this step, the received signal, $\rcvYslotNoised{t}$, is composed of diffusion channel output and the noise term,
\beqn
 \rcvYslotNoised{t} = \rcvYslot{t} + w(t, t+\ts)
 \label{eqn:received_y_slot_wtNoise}
\eeqn
where $w(t,t+\ts)$ follows Gaussian distribution $\mathcal{N}(0,\sigma_w^2)$ for each component independently. 

Demodulator module gets the $\rcvYslotNoised{t}$ and tries to decode the intended symbol. It also has connection with ISI mitigation module and removes the ISI effect of the previous symbol depending on the previous decision. If we use both (\ref{eqn:signal_tap_model}) and (\ref{eqn:received_y_slot_wtNoise}), we can remove the effect of the previous symbol at the receiver side as shown in (\ref{eqn:filteringISI}).
\beqn
 \rcvYslotNoised{t} - H_1 \sentXpslot{t} = H_0 \sentXslot{t} + w(t, t+\ts)
 \label{eqn:filteringISI}
\eeqn
Demodulator applies the symbol detection function, $\delta(.)$, on the filtered signal at the receiver side.

End-to-end simulator tracks all the released molecules for each symbol, hence, we see the ISI effect more than one symbol. For the simplicity demodulator module, however, just considers the previous decision for feedback. Even with just considering the one previous decision amplitude based demodulation performance increases significantly. 

%%%%%%%%%%%%%%%%%%%%%%%%%%%%%%%%%%%%%%%%%%%%%%%%%%%%%%%%%%%%%%%%%%%%%%%%%%%%%%%%
\begin{figure*}[!t]
\begin{center}
\subfigure[BCSK modulated on square wave]
{\includegraphics[width=0.48\textwidth,keepaspectratio]{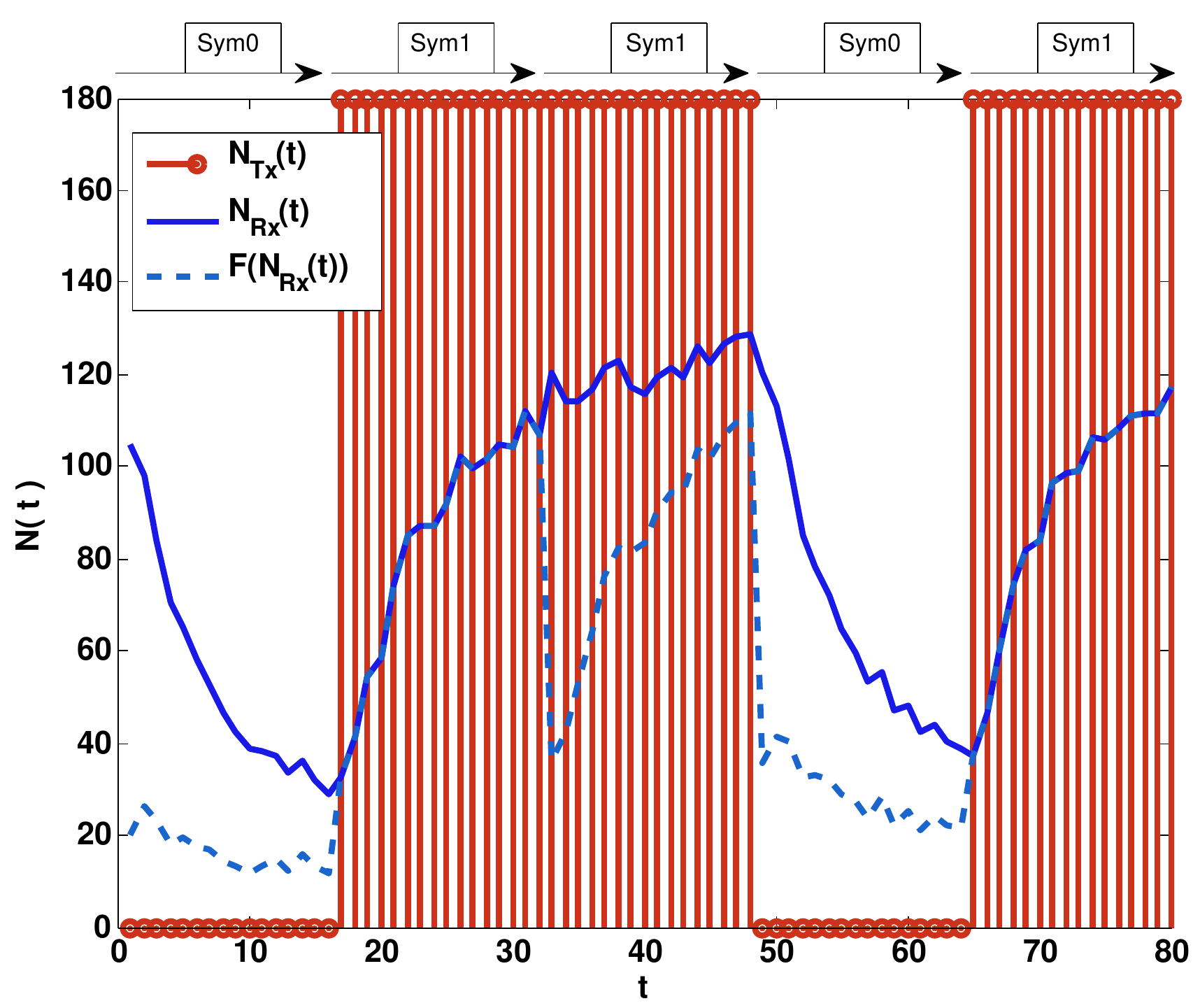}
\label{fig:sub1:bcsk} } \subfigure[BMFSK modulated on cosine wave]
{\includegraphics[width=0.48\textwidth,keepaspectratio]{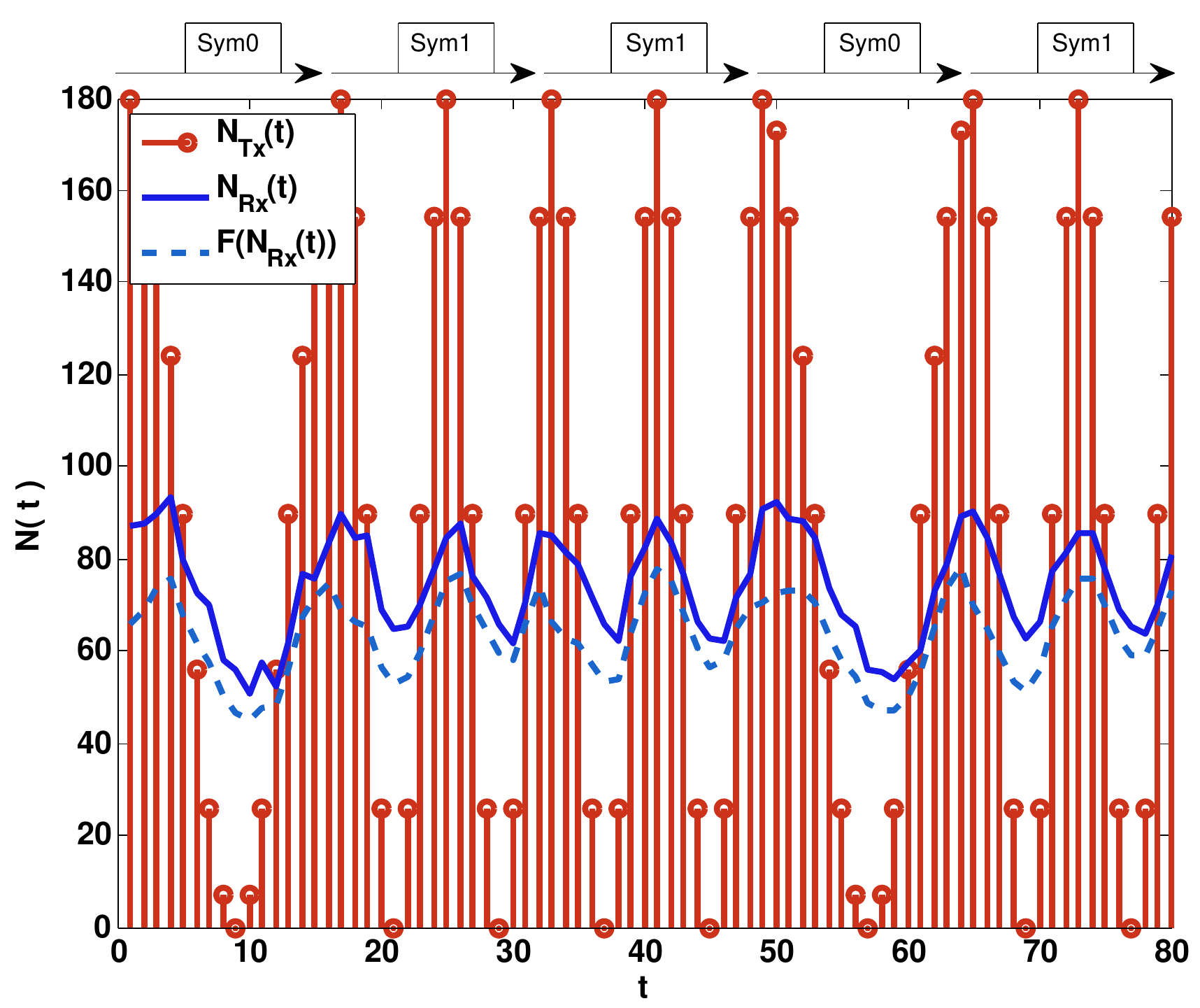}
\label{fig:sub1:bmfsk} }
\end{center}
\caption{Diffusion channel output signals for BCSK and BMFSK.}
\label{fig:diffusion_channel_output}
\end{figure*}
%%%%%%%%%%%%%%%%%%%%%%%%%%%%%%%%%%%%%%%%%%%%%%%%%%%%%%%%%%%%%%%%%%%%%%%%%%%%%%%%
\section{Performance Evaluation}
\label{Sec:performance_evaluation}
The communication environment of this work is constructed similar to the work in \cite{kuran2010energy}. Performance evaluation is carried out using the human insulin hormone as the messenger molecule and a device whose parameters are similar to a pancreatic $\beta$-cell as the transmitter \footnote{Other messenger molecules including isomers proposed in~\cite{TNBS_Kim13} are also possible.}. The propagation environment is chosen as water in the human body temperature. The size of the receiver unit is chosen as an average eukaryotic cell.  We developed an end-to-end simulator that randomly creates 2000 symbols to send consecutively via the CvD channel where the distance is $1 \mu m$. The parameters of the system are given in Table~\ref{tab:simparams} 

%%%%%%%%%%%%%%%%%%%%%%%%%%%%%%%%%%%%%%%%%%%%%%%%%%%%%%%%%%%%%%%%%%%%%%%%%%%%%%%%
\begin{table}[htb]
\caption{Simulation parameters.}
\label{tab:simparams}
\centering
\begin{tabular}{ll}
\hline  Parameter   &Value                                        \\
\hline  $D$                           &$79.4 \mu m^2/s$	 \\
        $r_{TN}$, $r_{RN}$        &$10 \mu m$             \\
        $r_{MM}$                	   &$2.5 nm$                  \\
        $d$                               &$1 \mu m$               \\
        $\ts$                            &$0.032 s$                   \\
        $\tss$                            &$0.002 s$                   \\
        $\numsym$                  &$2000$                     \\
        $\meanAmp$                 &$90$ molecules        \\
        Simulation repetition    &$20$                          \\
%        MFSK Periods                  &$\{0.32, 0.16, 0.1, 0.05\}$ s\\
\hline
\end{tabular}
\end{table}
%%%%%%%%%%%%%%%%%%%%%%%%%%%%%%%%%%%%%%%%%%%%%%%%%%%%%%%%%%%%%%%%%%%%%%%%%%%%%%%%

We first give the output signal of diffusion channel for all the modulations considered. Then we give Receiver Operating Characteristics (ROC) curves and Symbol Error Rate (SER) values for performance analysis. SNR value is defined as follows
\beqn
 SNR[dB] = 10 \log_{10} \left(\frac{\meanAmpRx}{\sigma_w}\right)^2 
\eeqn
where $\meanAmpRx$ and $\sigma_w$ are mean amplitude of the received signal and the standard deviation of the noise mentioned in (\ref{eqn:received_y_slot_wtNoise}), respectively. We have mean transmitted and received power same/similar for all of the modulation schemes. Even if we do not know the original noise floor in the environment we added different noise levels in the system and get the SNR for having SNR term analogous to classical communication systems. Performance metric shows us the achievable performance under different hypothetical noise levels.

\subsection{Received Signal, $\rcvYt$, Analysis}
In Fig.~\ref{fig:diffusion_channel_output}, $\sentXt$, $\rcvYt$, and filtered $\rcvYt$ signals for five consecutive symbols are presented for two different modulation schemes. Filtered $\rcvYt$ signal is represented by $F(\rcvYt)$ in the figures. Noise term is not included in the received signal for understanding the effect of diffusion channel only. 

In the BCSK case, the ISI filter removes the effect of one previous symbol, hence, the symbol shapes become similar for ``0" and ``1" separately.  For the symbol ``0", signal shape becomes like a horizontal line and for the symbol ``1", it starts from the level of the symbol ``0" and rises significantly.  Therefore, detecting the amplitude-based modulations becomes easier after applying the ISI mitigation. Otherwise, the gap between two symbols in terms of sum of $\rcvYt$ is not enough to determine the intended symbol. If we do not apply the ISI filter, sum of $\rcvYt$ for symbol ``1" while rising to a level and for symbol ``0" while coming back to rest position are similar. Hence, ISI mitigation is necessary for efficient symbol demodulation for amplitude-based modulations.

In the BMFSK case, ISI filter is applied with a coefficient not to affect the periodic properties of the signal too much. We observe that the received signal catches the peaks and it is possible to demodulate the symbols effectively for BMFSK case. Since $\rcvYt$ catches the peaks the correlation detector can determine the intended symbol effectively. Periodicity of the symbol is not degraded for low frequencies, hence ISI is not effecting the symbol demodulation primarily. Therefore, ISI filter helps amplitude-based modulations more effectively.  

%%%%%%%%%%%%%%%%%%%%%%%%%%%%%%%%%%%%%%%%%%%%%%%%%%%%%%%%%%%%%%%%%%%%%%%%%%%%%%%%
\begin{figure}[!t]
\centering
\includegraphics[width=0.99\columnwidth,keepaspectratio] {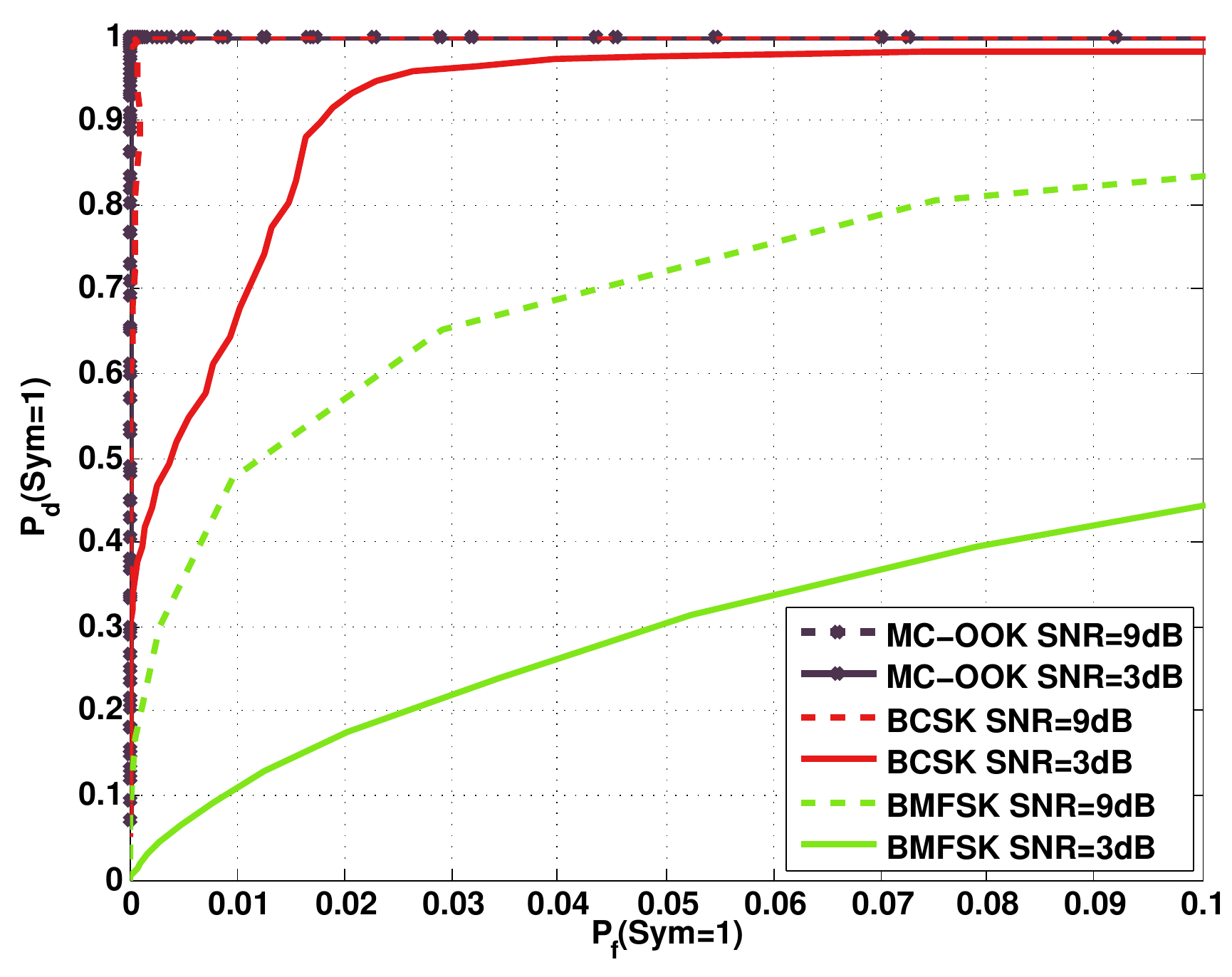}
\caption{ROC curves for symbol ``1" of binary modulations.}
\label{fig:roc1_curves_binary_mod}
\end{figure}
%%%%%%%%%%%%%%%%%%%%%%%%%%%%%%%%%%%%%%%%%%%%%%%%%%%%%%%%%%%%%%%%%%%%%%%%%%%%%%%%

\subsection{Receiver Operating Characteristics (ROC) Analysis }
For analyzing symbol detection and false alarm probabilities we compare ROC curves. In Fig.~\ref{fig:roc1_curves_binary_mod}, ROC curves for symbol ``1" of binary modulations are depicted for 3 and 9 dB SNR values. $\pdsym{1}$ represents the probability of demodulating the received symbol as ``1" when the transmitted symbol is actually the symbol ``1". Similarly,  $\pfsym{1}$ corresponds to the probability of demodulating the received symbol as ``1" when the transmitted symbol is actually the symbol ``0", hence it can be seen as false alarm of saying demodulated symbol is the symbol ``1".

In Fig.~\ref{fig:roc1_curves_binary_mod}, for a given false alarm constraint achievable detection probability can be compared. For $\pfsym{1}=0.05$, we have nearly 0.3, 0.98, and 1 as a detection probability for symbol ``1" for 3 dB case for BMFSK, BCSK, and MC-OOK, respectively. Increasing SNR results in better detection probability. Hence, with the same false alarm constraint we have 0.72, 1, and 1 as a detection probability for symbol ``1" for 9 dB case for the same order of modulations. MC-OOK has the best performance in terms of $\pdsym{1}$ with given $\pfsym{1}$ constraint. BCSK has comparable performance for 9 dB SNR while 3 dB case can also be satisfactory for some cases.

\subsection{SER Analysis}
\begin{figure}[!t]
\centering
\includegraphics[width=0.99\columnwidth,keepaspectratio] {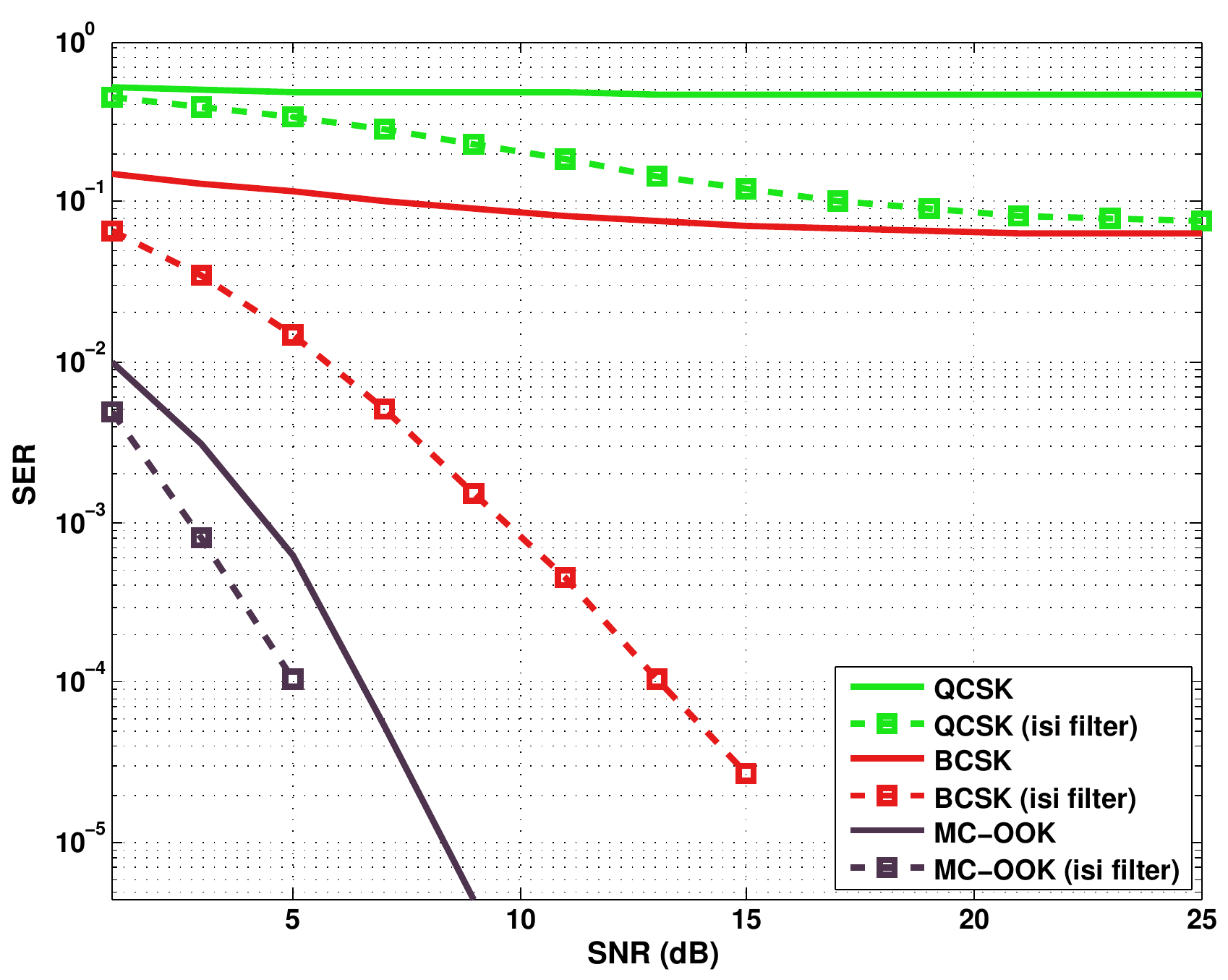}
\caption{SER plots for amplitude-based modulations.}
\label{fig:ser_plots_amplitude}
\end{figure}

In Fig.~\ref{fig:ser_plots_amplitude}, SNR versus SER values for amplitude-based modulations are shown. In the figure x-axis corresponds to SNR and the y-axis corresponds to SER. When SNR increases the error rate decreases and MC-OOK is the best performing modulation scheme. MC-OOK, however, has the highest $\pammrtx$ value which is 32 times higher than the alternative modulations for the parameters given before. 
%%%%%%%%%%%%%%%%%%%%%%%%%%%%%%%%%%%%%%%%%%%%%%%%%%%%%%%%%%%%%%%%%%%%%%%%%%%%%%%%
%%%%%%%%%%%%%%%%%%%%%%%%%%%%%%%%%%%%%%%%%%%%%%%%%%%%%%%%%%%%%%%%%%%%%%%%%%%%%%%%

Curves with the square markers are the demodulators using ISI filter. Results show that having an ISI filter improves the error rate, with mitigating the effect of the one previous symbol. Since diffusion channel spreads the signal to a very long duration variance of the hitting times is high. Hence, QCSK both with and without ISI filter and BCSK without ISI filter have an error floor at $0.07$, $0.4$, and $0.06$, respectively. BCSK error floor at $0.06$ can be mitigated via employing ISI filter that considers one previously demodulated symbol decision feedback.

\begin{figure}[!t]
\centering
\includegraphics[width=0.99\columnwidth,keepaspectratio] {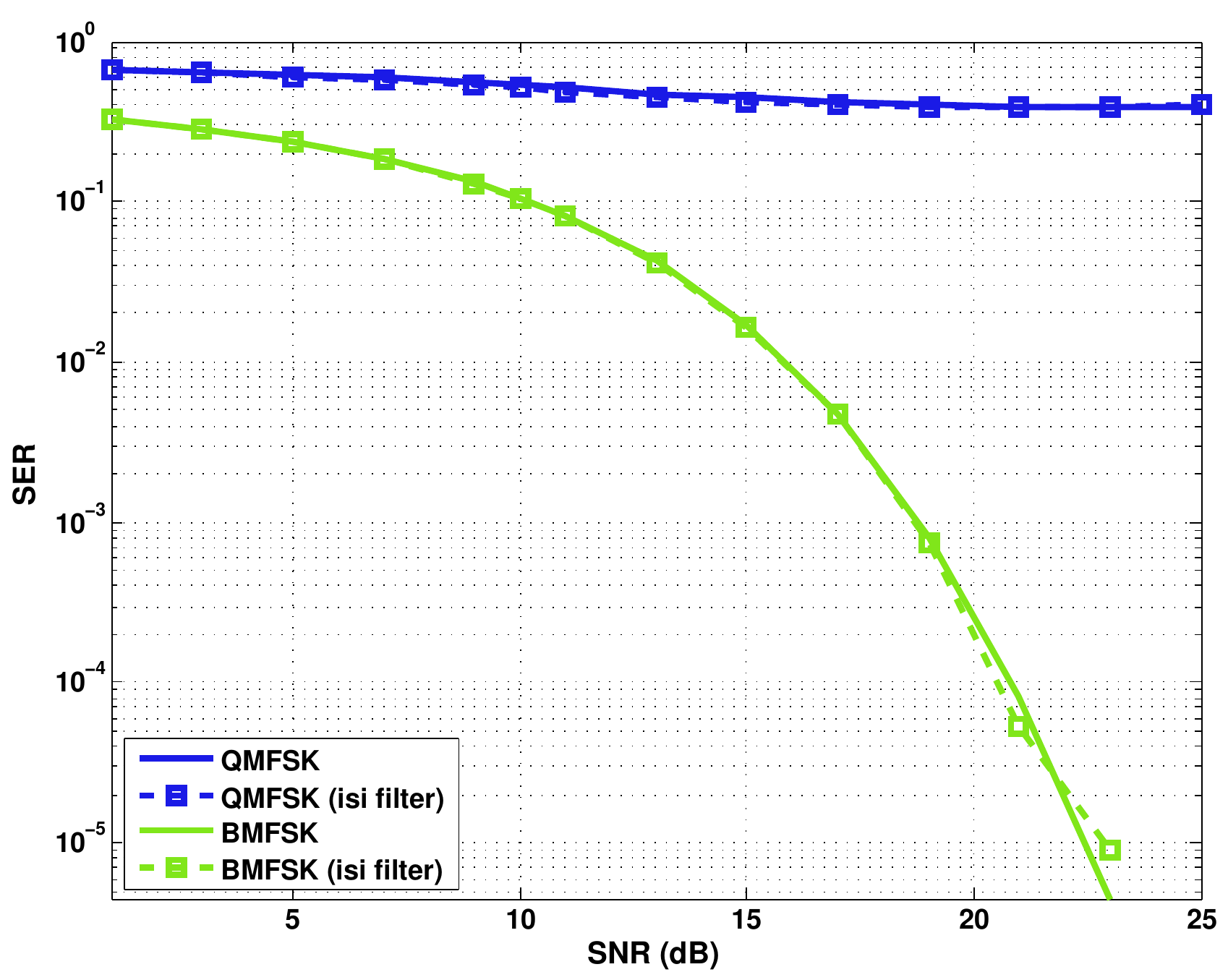}
\caption{SER plots for frequency-based modulations.}
\label{fig:ser_plots_frequency}
\end{figure}

In Fig.~\ref{fig:ser_plots_frequency}, SNR versus SER values for frequency-based modulations are shown. When SNR increases the error rate decreases significantly for BMFSK, however, for QMFSK system has an error floor which could not be mitigated. Curves with the square markers are the demodulators using ISI filter and we see that having an ISI filter does not affect the system significantly. 
%%%%%%%%%%%%%%%%%%%%%%%%%%%%%%%%%%%%%%%%%%%%%%%%%%%%%%%%%%%%%%%%%%%%%%%%%%%%%%%%
%%%%%%%%%%%%%%%%%%%%%%%%%%%%%%%%%%%%%%%%%%%%%%%%%%%%%%%%%%%%%%%%%%%%%%%%%%%%%%%%
%%%%%%%%%%%%%%%%%%%%%%%%%%%%%%%%%%%%%%%%%%%%%%%%%%%%%%%%%%%%%%%%%%%%%%%%%%%%%%%%
\begin{figure}[!t]
\centering
\includegraphics[width=0.99\columnwidth,keepaspectratio] {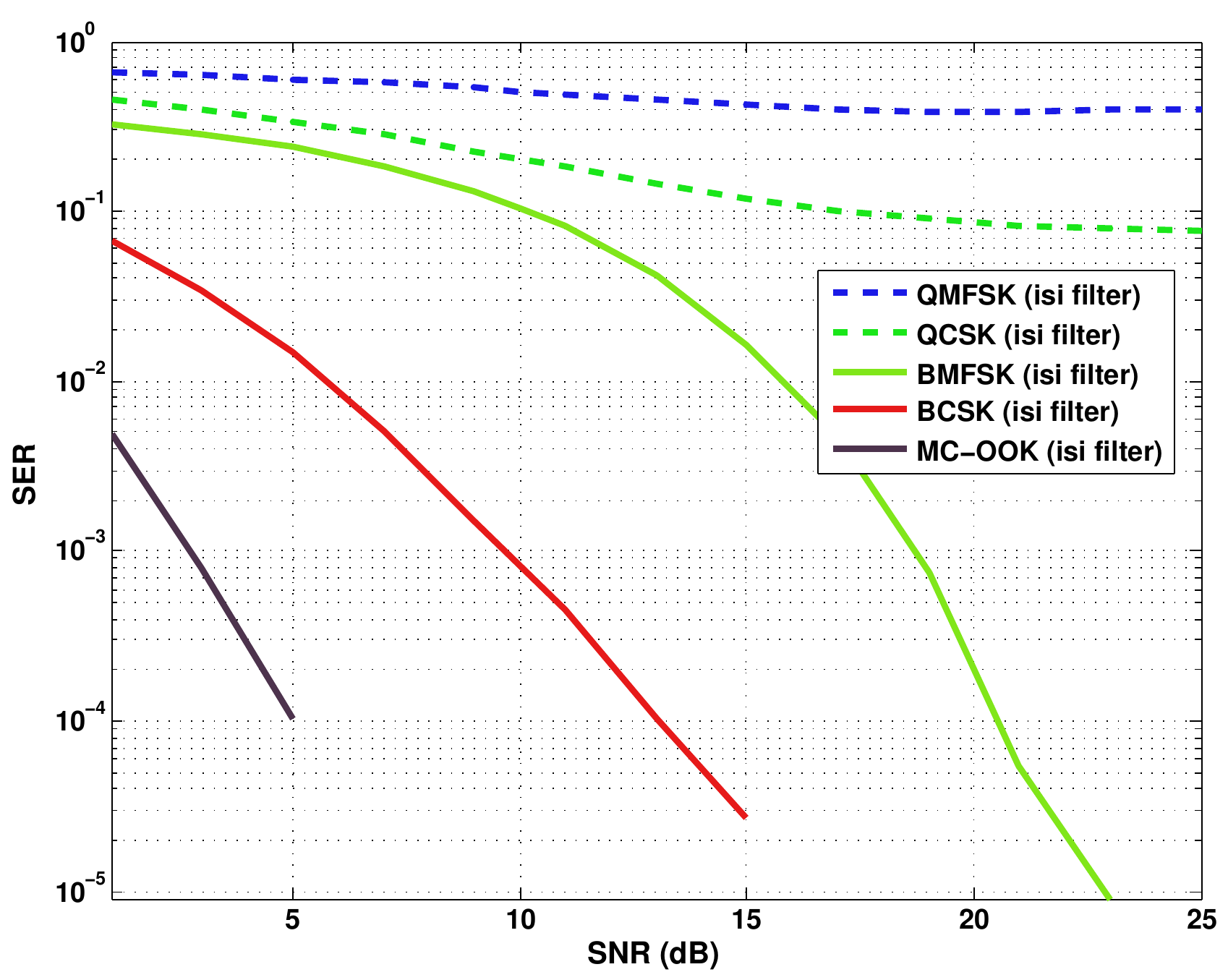}
\caption{SER plots for modulations with ISI filtering.}
\label{fig:ser_plots_all_isi_filtering}
\end{figure}
%%%%%%%%%%%%%%%%%%%%%%%%%%%%%%%%%%%%%%%%%%%%%%%%%%%%%%%%%%%%%%%%%%%%%%%%%%%%%%%%

%%%%%%%%%%%%%%%%%%%%%%%%%%%%%%%%%%%%%%%%%%%%%%%%%%%%%%%%%%%%%%%%%%%%%%%%%%%%%%%%
\begin{table*}[htb]
\caption{Modulation comparison matrix.}
\label{tab:comparison}
\centering
\begin{tabular}{llllll}
\hline  Metric    & MC-OOK   & BCSK     &   BMFSK &   QCSK &    QMFSK   \\
\hline  
        SER       & Low     & Moderate & Moderate &  High  &  High	 \\
        $\pammrtx$ & High    & Low      & Low      & Low    & Low       \\
        Node Complexity & Low    & Moderate      & High      & Moderate    & High       \\
\hline
\end{tabular}
\end{table*}
%%%%%%%%%%%%%%%%%%%%%%%%%%%%%%%%%%%%%%%%%%%%%%%%%%%%%%%%%%%%%%%%%%%%%%%%%%%%%%%%

In Fig.~\ref{fig:ser_plots_all_isi_filtering}, all the modulations with ISI filter is considered and the results are depicted. Incresing SNR results in decrease in SER, however,  in quadrature modulations there are error floors and more advanced techniques are required to mitigate these error floors. MC-OOK has the best performance compared to the BCSK modulated on a square wave and BMFSK modulated on a cosine wave. MC-OOK, however, has the highest $\pammrtx$ value (32 times higher compared to other modulations) while the BCSK has an acceptable SER and $\pammrtx$ value.  

In Table~\ref{tab:comparison}, we summarize the performance and the capabilities of the modulation schemes. In terms of SER MC-OOK performs better than the other schemes, however, in terms of $\pammrtx$ it is the far worse than the others. These results and the summary table can help to the system designers in nanonetworking domain.

\section{Conclusions}
\label{Sec:conclusion}
In this paper, we developed a 3-dimensional molecular communication simulator to track all the transmitted messenger molecules precisely and to investigate the effect of ISI. To eliminate the ISI effect, a simple two-tap decision feedback equalizer filter is applied, and several types of modulation techniques were analyzed to evaluate the system performance. From the simulator, MC-OOK showed the best performance in terms of detection probability with a fixed false alarm constraint. This technique, however, has the highest $\pammrtx$ compared to the CSK and FSK. In case of constraints on transmitter $\pammrtx$ value due to storage volume and synthesis process, it may not be possible to utilize MC-OOK. Additionally, BMFSK generates highly decreasing SER with increasing SNR whereas quadrature technique has a certain error floor to be mitigated. 

For the future work, the filter module can be advanced to decrease the SER even in FSK modulation. Moreover, We can revisit $\pammrtx$ to define more realistic optimization problems.

%ACKNOWLEDGMENTS are optional
\section{Acknowledgments}
This research is funded by the MSIP (Ministry of Science, ICT \& Future Planning), Korea, under the ``IT Consilience Creative Program" (NIPA- 2014-H0201-14-1001) supervised by the NIPA (National IT Industry Promotion Agency).

\bibliographystyle{IEEEtran}
\bibliography{IEEEabrv,sigproc}  

\end{document}